\documentclass[10pt,conference]{IEEEtran}
\IEEEoverridecommandlockouts
\makeatletter

\usepackage[colorinlistoftodos,prependcaption,textsize=tiny]{todonotes}
\usepackage{regexpatch}
\xpatchcmd{\@todo}{\setkeys{todonotes}{#1}}{\setkeys{todonotes}{inline,#1}}{}{}

\usepackage{comment}
\let\wfs@comment@comment\comment
\let\comment\@undefined
\usepackage{changes}
\let\wfs@changes@comment\comment
\let\comment\@undefined
\newcommand\comment{%
    \ifthenelse{\equal{\@currenvir}{comment}}
    {\wfs@comment@comment}
    {\wfs@changes@comment}%
}
\makeatother

\usepackage{amsmath}
\usepackage{hyperref}
\usepackage{enumitem}
\usepackage{multirow}
\usepackage{outlines}
\usepackage{graphicx}
\usepackage{algorithm}
\usepackage{float}
\usepackage{nicefrac}
\usepackage{url}
\usepackage{booktabs}
\usepackage{listings}
\definechangesauthor[name="Ilya Tyagin", color=teal]{it}

\usepackage{cite}
\usepackage{url}
\usepackage{amsmath,amssymb,amsfonts}
\usepackage{algorithmicx}
\usepackage{algorithm}
\usepackage{algpseudocode}
\usepackage{graphicx}
\usepackage{physics}
\usepackage{textcomp}
\usepackage{hyperref}
\usepackage{xcolor} 
\usepackage{booktabs}

\definechangesauthor[color=red]{is}
\definechangesauthor[color=purple]{KS}
\newcommand{\sysname}{QAOA-GPT}

\def\BibTeX{{\rm B\kern-.05em{\sc i\kern-.025em b}\kern-.08em
    T\kern-.1667em\lower.7ex\hbox{E}\kern-.125emX}}
\begin{document}
\title{QAOA-GPT: Efficient Generation of Adaptive and Regular Quantum Approximate Optimization Algorithm Circuits}%

\author{\IEEEauthorblockN{Ilya Tyagin}
\IEEEauthorblockA{Department of Computer and Information Sciences \\
University of Delaware\\
Newark DE, USA \\
tyagin@udel.edu}
\and
\IEEEauthorblockN{Marwa H. Farag}
\IEEEauthorblockA{NVIDIA Corporation\\
Santa Clara CA, USA \\
mfarag@nvidia.com}
\and
\IEEEauthorblockN{Kyle Sherbert}
\IEEEauthorblockA{Department of Chemistry\\
Department of Physics\\
Center for Quantum Information Science\\ and Engineering\\
Virginia Tech \\
Blacksburg VA, USA \\
kyle.sherbert@vt.edu}
\and
\IEEEauthorblockN{Karunya Shirali}
\IEEEauthorblockA{Department of Physics\\
Center for Quantum Information Science\\ and Engineering\\
Virginia Tech \\
Blacksburg VA, USA \\
karunyashirali@vt.edu}
\and
\IEEEauthorblockN{Yuri Alexeev}
\IEEEauthorblockA{NVIDIA Corporation\\
Santa Clara CA, USA \\
yalexeev@nvidia.com}
\and
\IEEEauthorblockN{Ilya Safro}
\IEEEauthorblockA{Department of Computer and Information Sciences \\
Department of Physics and Astronomy \\
University of Delaware\\
Newark DE, USA \\
isafro@udel.edu}
}

\maketitle

\begin{abstract}
    
    Quantum computing has the potential to improve our ability to solve certain optimization problems that are computationally difficult for classical computers, by offering new algorithmic approaches that may provide speedups under specific conditions. 
    In this work, we introduce QAOA-GPT, a generative framework that leverages Generative Pretrained  Transformers (GPT) to directly synthesize quantum circuits for solving quadratic unconstrained binary optimization problems, and demonstrate it on the MaxCut problem on graphs. 
    To diversify the training circuits and ensure their quality, we have generated a synthetic dataset using the adaptive QAOA approach, a method that incrementally builds and optimizes problem-specific circuits. 
    The experiments conducted on a curated set of graph instances demonstrate that QAOA-GPT, generates high quality quantum circuits for new problem instances unseen in the training as well as successfully parametrizes QAOA. 
    Our results show that using QAOA-GPT to generate quantum circuits will significantly decrease both the computational overhead of classical QAOA and adaptive approaches that often use gradient evaluation  to generate the circuit and the classical optimization of the circuit parameters. 
    Our work shows that generative AI could be a promising avenue to generate compact quantum circuits in a scalable way.
    
\end{abstract}


\section{Introduction}

Quantum computing is rapidly emerging technology with significant potential across various domains, including finance \cite{herman2023quantum}, chemical simulations \cite{cao2019quantum}, material science \cite{matsci_qc}, combinatorial optimization \cite{shaydulin2019hybrid}, and machine learning \cite{biamonte2017quantum}, among others.
Variational quantum-classical algorithms represent one of the most promising classes of quantum algorithms in different domains, showing potential for both fault-tolerant quantum computers and near-term noisy intermediate-scale quantum (NISQ) devices. The Quantum Approximate Optimization Algorithm (QAOA) \cite{farhi2014QAOA} and many of its subsequent versions and customizations \cite{blekos2024review} belong to this class and demonstrate great potential due to their problem/application flexibility and compatibility with various quantum architectures.

The original QAOA framework employs a fixed ansatz structure, which can limit expressibility and hinder performance, particularly on near-term quantum devices where circuit depth is limited. This rigid design may not capture the problem-specific features needed for efficient optimization. Such methods as ADAPT-QAOA \cite{zhu2022adaptive} %
address this challenge by iteratively constructing the ansatz in a problem-informed manner. At each step, ADAPT-QAOA selects operators from a predefined pool based on their gradient with respect to the cost function, incorporating only those that contribute most significantly to improving the objective. This adaptive strategy often leads to more compact and expressive circuits, making it especially suitable for both NISQ and future fault-tolerant quantum architectures. Similar adaptive strategies have been developed for variational quantum eigensolvers \cite{grimsley2019adaptive} and combinatorial optimization \cite{liu2022layer} to mention just a few. 
Despite their  advantages, ADAPT-QAOA and other similar methods have notable limitations. The iterative ansatz construction requires repeated gradient evaluations, which can be computationally expensive and slow down the overall optimization process. Additionally, the resulting circuits may have irregular structures that are not well-suited for current quantum hardware, particularly those with limited qubit connectivity or  gate constraints.

Recent advances in generative artificial intelligence (AI), particularly in large-scale models like Generative Pretrained Transformers (GPT), have opened new avenues for automating algorithm design and code generation. In the quantum domain, recent works such as Grover-GPT~\cite{wang2024grovergpt} and the Quantum Generative Eigensolver (GQE)~\cite{nakaji2024generative} have demonstrated the potential of AI to assist in quantum algorithm discovery. These models can extract complex patterns and generate structured outputs, making them natural candidates for tasks involving sequential quantum circuit construction.

In this work, we propose a novel GPT-based generative framework to overcome the scalability limitations of QAOA and some of its subsequent versions as well as the slowness of adaptive methods. Instead of relying on slow, iterative, layer-by-layer optimization, we train a GPT model to directly generate efficient quantum circuits tailored to specific graph-based optimization problems. 
Our model operates over generalized Pauli strings, allowing hardware-agnostic circuit representations while remaining extensible to future hardware-specific implementations.

This AI-driven approach offers multiple advantages:
\begin{enumerate}
    \item Significantly reduced circuit generation time through inference rather than optimization,
    \item Improved scalability as sequence length grows,
    \item Efficient use of GPU-accelerated and distributed architectures, and
    \item Potential discovery of latent structural patterns in quantum algorithms that are not accessible through classical heuristics.
\end{enumerate}

\noindent {\bf Our contribution}: 
We introduce a GPT-based generative model \sysname~ that directly synthesizes problem-specific quantum circuits, bypassing the need for iterative optimization of variational algorithms (e.g., ADAPT-QAOA that is used for the training set generation, i.e., the preprocessing stage). Our method addresses key scalability bottlenecks in variational quantum algorithms and establishes a foundation for integrating large-scale generative AI into quantum algorithm design.  The generated circuits maintain high solution quality across a broad range of graph instances of various problem sizes and require only a single forward pass for inference, enabling orders-of-magnitude faster circuit generation compared to iterative methods. QAOA-GPT has been trained on the synthetic dataset generated by the ADAPT-QAOA implementation in two codes: Julia and the CUDA-Q platform~\cite{bayraktar2023cuquantum, cudaq_github}. This work serves as a proof-of-concept demonstrating the feasibility and performance advantages of AI-assisted quantum circuit generation, bridging the gap between classical optimization-based methods and scalable, data-driven approaches.

\section{Background}
\subsection{MaxCut as a Combinatorial Optimization Problem}

The \textit{MaxCut problem} is a well-known combinatorial optimization problem that often serves as a benchmark for various classical and quantum optimization algorithms. It is classified as \textit{NP-hard}~\cite{NP_completeQAOA}, implying that no known deterministic polynomial-time algorithm can solve all instances of the problem efficiently.

Given a graph \( G = (V, E, w) \), where:
\begin{itemize}
    \item \( V \) is the set of vertices, $|V|=n$,
    \item \( E \) is the set of edges, and
    \item \( w: E \rightarrow \mathbb{R}^+ \) assigns weights to edges,
\end{itemize}
the goal of MaxCut is to partition the vertex set \( V \) into two disjoint subsets \( V_1 \) and \( V_2 \) such that the total weight of edges crossing the cut (i.e., connecting nodes from different subsets) is maximized.

This optimization task can be formulated as a \textit{Quadratic Unconstrained Binary Optimization (QUBO)} problem. Assigning each vertex \( i \in V \) a binary variable \( x_i \in \{-1, 1\} \), indicating its subset membership, the MaxCut objective function is:
\begin{equation}
    \label{eq: MaxCut_objective}
    \max_{x\in \{-1, 1\}^n} \sum_{(i,j) \in E} w_{ij}\frac{(1 - x_i x_j)}{2}
\end{equation}

This classical objective can be translated into a quantum Hamiltonian as \( - \frac{1}{2}\sum_{(i,j) \in E} w_{ij} (I - Z_i Z_j) \), where \( Z_i, Z_j \) are the Pauli-Z operator acting on qubits \( i, j \), respectively. This transformation yields a cost Hamiltonian \( H_c \) that can be used in quantum optimization algorithms such as QAOA. Many combinatorial optimization problems (including the Ising model) can be casted as QUBO and MaxCut \cite{glover2022quantum}.

\subsection{Quantum Approximate Optimization Algorithm (QAOA)}

The QAOA \cite{farhi2014QAOA} is a hybrid quantum-classical variational algorithm often used to solve combinatorial optimization problems by approximating the optimal solution via quantum state evolution and classical parameter optimization.

The QAOA circuit alternates between two types of Hamiltonians:
\begin{itemize}
    \item \textit{Cost Hamiltonian} $H_c$, derived from the objective function of the problem (e.g., the MaxCut formulation),
    \item \textit{Mixing Hamiltonian} \( H_B = \sum_{i=1}^n X_i \), where \( X_i \) is the Pauli-X operator.
\end{itemize}

The algorithm begins with the uniform superposition state \( \ket{+}^{\otimes n} \) and applies \( p \) layers of alternating unitary operations, namely, the cost and mixing unitaries, defined as 
    \[
    U_P(\gamma_i) = e^{-i \gamma_i H_c} \text{ and }
    U_M(\beta_i) = e^{-i \beta_i H_B}\text{, respectively.}
    \]

After \( p \) layers, the quantum state prepared by QAOA is:
\begin{equation}
    \label{eq: QAOA_state}
    \ket{\gamma, \beta} = U_{M}(\beta_p)U_{P}(\gamma_p)\cdots U_{M}(\beta_1)U_{P}(\gamma_1)\ket{+}^{\otimes n}
\end{equation}
where \( \gamma = \{\gamma_1, \ldots, \gamma_p\} \) and \( \beta = \{\beta_1, \ldots, \beta_p\} \) are sets of variational parameters.

The expectation value of the cost Hamiltonian is evaluated as
\[
\langle H_c \rangle = \bra{\gamma,\beta}H_c\ket{\gamma,\beta}.
\]
A classical optimizer is used to iteratively adjust \( \gamma \) and \( \beta \) to minimize (or maximize, depending on formulation) this expectation value. The parameters that yield the extremal value approximate a solution to the original optimization problem.

QAOA was originally demonstrated on the MaxCut problem~\cite{farhi2014QAOA}. However, since it can also be directly casted as the general QUBO problem \cite{glover2018tutorial}, there are many other applications, e.g., graph partitioning and community detection \cite{ushijima2021multilevel}, graph vertex \cite{bravyi2022hybrid} and edge \cite{tsvelikhovskiy2024equivariant} coloring, and portfolio optimization \cite{herman2023quantum}.

\subsection{ADAPT-QAOA}
\label{sec:adapt_qaoa}
An important advancement in variational quantum algorithms is the \textit{Adaptive Derivative-Assembled Problem-Tailored Quantum Approximate Optimization Algorithm} (ADAPT-QAOA) proposed by Zhu \textit{et al.}~\cite{zhu2022adaptive}. ADAPT-QAOA extends the standard QAOA by replacing the fixed mixer Hamiltonian with a dynamically constructed, problem-tailored ansatz that is iteratively built from an operator pool. Inspired by the ADAPT-VQE framework~\cite{grimsley2019adaptive}, this approach addresses known limitations of standard QAOA such as slow convergence and increased classical optimization overhead at high parameter counts.

In ADAPT-QAOA, the quantum ansatz is grown one layer at a time using a gradient-based selection criterion. At each iteration $k$, operators $A_j$ from a predefined mixer pool are evaluated via the energy gradient
\[
g_j = \frac{\partial E}{\partial \beta_j} = -i \left\langle \psi^{(k-1)} \left| e^{i \gamma_0 H_c} [H_c, A_j] e^{-i \gamma_0 H_c} \right| \psi^{(k-1)} \right\rangle,
\]
where $H_c$ is the cost Hamiltonian, $\gamma_0$ is the initial variational parameter of the problem Hamiltonian  and $g_j$ is the energy ($E$) gradient with respect to the operator pool. The operator $A^{(k)}$ with the largest gradient norm is selected and appended to the circuit as $e^{-i \beta_k A^{(k)}} e^{-i \gamma_k H_c}$, producing the updated variational state:
\[
|\psi^{(k)}\rangle = e^{-i \beta_k A^{(k)}} e^{-i \gamma_k H_c} |\psi^{(k-1)}\rangle.
\]
based on the previous state $|\psi^{(k-1)}\rangle$.
The mixer pool may include single-qubit Pauli operators ($X_i$, $Y_i$), their sums, or two-qubit entangling terms (e.g., $X_i Z_j$, $Y_i Y_j$), constrained to commute with the $\mathbb{Z}_2$ symmetry generator $F = \bigotimes_i X_i$ in order to preserve the symmetry of Ising-type Hamiltonians.

After each layer is added, all variational parameters $\{\beta_m, \gamma_m\}_{m=1}^k$ are re-optimized to minimize the cost function $\langle \psi^{(k)} | H_c | \psi^{(k)} \rangle$. The iterative process continues until the gradient norm falls below a predefined threshold. For more details we refer the reader to \cite{zhu2022adaptive}.

\subsection{Generative Pre-trained Transformer (GPT)}
\label{sec:gpt}

The Generative Pre-trained Transformer (GPT) is a semi-supervised learning framework that leverages large-scale unlabeled text to learn transferable language representations via generative pre-training \cite{radford2018improving}. The core methodology employs a \textit{decoder-only transformer architecture} \cite{vaswani2017attention}, composed of masked self-attention layers that enable autoregressive modeling of sequential text data.

In the pre-training phase, GPT optimizes a \textit{left-to-right language modeling objective} over a corpus of contiguous text snippets. Given a sequence of tokens $U = \{u_1, u_2, \dots, u_n\}$, the model is trained to maximize the log-likelihood of each token conditioned on its preceding context:

\begin{equation}
\label{eq: gpt_loss}
    \mathcal{L}_{\text{LM}}(U) = \sum_{i=1}^{n} \log P(u_i \mid u_1, u_2, \dots, u_{i-1}; \Theta),
\end{equation}
where $\Theta$ represents the parameters of the transformer model.

GPT uses multi-layer transformer blocks consisting of masked multi-head self-attention mechanisms and position-wise feed-forward networks. Inputs are embedded using a combination of \textit{token embeddings} and \textit{learned positional embeddings}, allowing the model to represent both the identity and position of tokens. The masking in the self-attention mechanism ensures that the prediction of token $u_i$ depends only on the previous tokens $\{u_1, \dots, u_{i-1}\}$, which aligns with the causal structure of natural language.

While we are not directly focused on the natural language capabilities of generative models like GPT, their token-based generation paradigm is particularly well-suited for source code synthesis tasks \cite{chen2021evaluating}. Quantum circuits, which can be naturally represented as sequences of unitary operations, align well with this framework. As such, we leverage GPT's ability to model and generate structured sequences in order to discover potentially more efficient quantum circuits. Our interest lies in exploring GPT's capacity to scale this approach effectively, given access to a sufficiently rich training dataset in the combinatorial optimization domain.

\subsection{Graph-level embeddings (FEATHER)}
\label{sec:feather}

FEATHER~\cite{rozemberczki2020characteristic} is a non-parametric graph embedding method based on \textit{characteristic functions} that represent local distributions of node features. For each node $u$, the method computes the \textit{r-scale random walk weighted characteristic function}:

\[
\phi_u(\theta, r) = \sum_{w \in V} \hat{A}^r_{u,w} \cdot e^{i\theta x_w}
\]
where $\hat{A} = D^{-1}A$ is the normalized adjacency matrix and $\hat{A}^r_{u,w}$ is the probability of reaching node $w$ from $u$ in $r$ steps via a random walk. This function is evaluated at a set of $d$ frequencies $\theta \in \Theta$ to form a complex-valued embedding that captures the higher-order structure and attribute distribution around each node.

Graph-level representations are constructed by pooling node-level embeddings (e.g., using the mean), resulting in descriptors that are invariant to graph isomorphisms and robust to localized feature noise. FEATHER supports multiple input features and scales linearly with the graph size, providing an efficient alternative to kernel or subgraph-based models.

In the context of this work, we need graph embedding to represent structural features of the graph and use them in the GPT training and inference stages.

\section{\sysname ~Method}
\label{sec:proposed_method}

\begin{figure}
    \centering
    \includegraphics[
        width=1\linewidth,
    ]{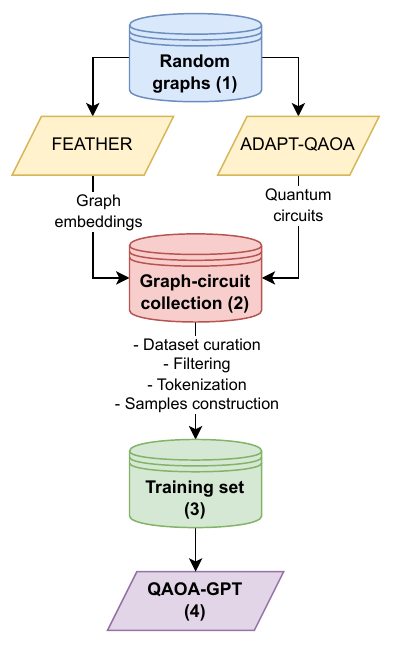}
    \caption{
      \label{fig:training_diagram}
      Schematic representation of the QAOA-GPT framework. Random graphs are independently processed by FEATHER to extract graph embeddings and by ADAPT-QAOA to generate optimized quantum circuits. These components form a curated graph–circuit collection, which is then tokenized and transformed into a structured training set used to train the QAOA-GPT model.
    }
\end{figure}

We propose a decoder-only transformer model (\sysname) trained to generate quantum circuits for solving weighted MaxCut problems. Figure~\ref{fig:training_diagram} outlines the full training pipeline, which consists of four main stages: (1) random graphs generation for training set, (2) graph–circuit dataset construction, (3) training set preparation, and (4) model training.

\subsection*{Step 1: Random Graph Generation}

We begin by sampling input graphs from the Erdős–Rényi distribution \( \mathcal{G}(n, s) \), where \( n \) denotes the number of nodes and \( s \) is the edge probability controlling graph sparsity. In this model, we restrict generation to connected graphs without isolated nodes or multiple components. Then, each edge $ij$ is assigned with a random weight $w_{ij}\in \mathcal{U}(0,1)$.

Each generated graph is subsequently processed by two components:
\begin{itemize}
    \item \textbf{Graph Embedding (FEATHER)}: computes (fixed-length) low dimensional vector representations based on characteristic functions of random walk distributions, used as conditioning input for the generative model.
    \item \textbf{Circuit Generation (ADAPT-QAOA)}: generates quantum circuits via adaptive ansatz construction, producing high-quality approximate solutions to the MaxCut problem.

    The ADAPT-QAOA circuits are constructed iteratively as described in ~\cite{zhu2022adaptive}:
\begin{enumerate}
    \item Initialize the  variational parameter \( \gamma_0 \) of the problem Hamiltonian.
    \item Select an operator \( O^{(k)} \in \mathcal{P} \) that has the largest gradient at each iteration \( k \) from a fixed operator pool base \( \mathcal{P} = \{O_1, O_2, \ldots, O_m\} \) 
    ~\cite{zhu2022adaptive}.  
    \item Append the operator and re-optimize all parameters \( \{ \gamma_1, \beta_1, \ldots, \gamma_k, \beta_k \} \in \mathbb{R}^{2k} \).
    \item Repeat the above steps until stopping criteria are achieved. The ADAPT-QAOA framework supports multiple stopping criteria that include the norm of the gradient, circuit depth and energy error-based conditions. In our work, only circuits achieving a target approximation ratio 
\[
\alpha = \frac{\langle \psi(\boldsymbol{\gamma, \beta}) | H_c | \psi(\boldsymbol{\gamma, \beta}) \rangle}{\text{OPT}(G)}
\]
are included, where \( H_c \) is the problem cost Hamiltonian derived from the graph and \( \text{OPT}(G) \) is the MaxCut value obtained via classical heuristics~\cite{BURER2002}.
\end{enumerate} 

To enrich the dataset, we vary \( \gamma_0 \) generating multiple valid circuits per graph and filter them later based on the the target $\alpha$.
\end{itemize}

\subsection*{Step 2: Graph–Circuit Collection}

Graph embeddings and ADAPT-QAOA circuits are stored together in a structured graph–circuit collection. 

\subsection*{Step 3: Training Set Construction}

The graph–circuit collection is curated and tokenized to form the final training set. Tokenization proceeds as follows:

\subsubsection*{Graph Tokens}

Each graph $G$ is represented by a weighted edge list:
\begin{align*}
\texttt{<bos>},\\
(i_1, j_1),\ w_{i_1j_1},\ \ldots,\ (i_{m}, j_{m}),\ w_{i_{m}j_{m}},\\
\texttt{<end\_of\_graph>}
\end{align*}
Here, \( i_k, j_k \in \{0, 1, \ldots, n-1\} \) denote node indices, \( (i_k, j_k) \in E \subseteq V \times V \), where $n$ is the number of nodes and $E$ is the set of edges in the graph $G$, and \( w_{ij} \in (0,1] \) is the weight of the edge between nodes \( i \) and \( j \).

\subsubsection*{Circuit Tokens}

Each circuit is expressed as a sequence of layer blocks:
\[
\texttt{<new\_layer\_p>},\ o_k,\ \gamma_k,\ \beta_k \quad \text{for } k = 1, \ldots, L
\]
where \( o_k \in \{0, \ldots, m-1\} \) is the operator index, and \( \gamma_k, \beta_k \in \mathbb{R} \) are the optimized parameters.

\subsubsection*{Numeric Handling}

All real numbers are rounded to two decimal places and clipped to \( [-10, 10] \). If any parameter exceeds this range, the corresponding circuit is not used.

\subsubsection*{Training Sample Construction}

Let $\mathcal{D} = \{d_1, \dots, d_N\}$ denote a dataset of $N$ independent token sequences, where each $d_i = [t_{i,1}, \dots, t_{i,L_i}]$ encodes a graph--circuit pair of length $L_i$. To ensure strict instance isolation, each $d_i$ is processed independently during training, without concatenation across sequences, unlike it is typically done for large-scale LLM training.

For each $d_i$, we sample a random window length $b_i \sim \mathrm{Uniform}(b_{\min}, b_{\max})$, where $b_{\max} \leq T$ and $T$ is the model’s context window length. Using a sliding window of size $b_i$, we extract subsequences
\[
w_i^{(k)} = [t_{i,k}, \dots, t_{i,k + b_i - 1}],
\]
and construct training pairs
\[
(x_i^{(k)}, y_i^{(k)}) = (w_i^{(k)}, w_i^{(k+1)}),
\]
corresponding to consecutive segments of each instance.

All sequences are then right-padded to length $T$, and attention masks are applied to pack sequences in batches. The main idea is that the model is never exposed to tokens from multiple graph-circuit instances within the same sequence, preserving the per-instance semantic boundary essential for structured circuit generation, which is different from standard language modeling approach.

\subsection*{Step 4: Model Architecture and Training \sysname}

The final training set is used to train \sysname, a decoder-only Transformer model based on the nanoGPT implementation of GPT-2. The model was trained from scratch and was not initialized from any pre-trained earlier model due to using custom circuit tokenization schema (i.e., we do not use fine tuning or prompt engineering).

Given a tokenized sequence \( \mathbf{x}_{1:T} \), the model uses:
\begin{itemize}
    \item Token embeddings \( \mathbf{E}_{\text{tok}} \in \mathbb{R}^{V \times d} \),
    \item Positional embeddings \( \mathbf{E}_{\text{pos}} \in \mathbb{R}^{T \times d} \),
    \item Graph embeddings \( \mathbf{e}_G \in \mathbb{R}^d \) from FEATHER.
\end{itemize}

The Transformer input is computed as:
\[
\mathbf{X} = \mathbf{E}_{\text{tok}}(\mathbf{x}_{1:T}) + \mathbf{E}_{\text{pos}} + \mathbf{e}_{G} \otimes \mathbf{1}_{1 \times T}
\]
where $T$ is the length of token sequence $X$, $V$ is the vocabulary size and \( \otimes \) denotes broadcasting the graph embedding across all tokens. This allows the model to condition the sequence generation on global graph features at every layer.

\subsection*{Training Procedure}

The model is trained via an autoregressive next-token prediction loss (Equation~\ref{eq: gpt_loss}), identical to standard GPT models~\cite{radford2018improving}. %
During the training, we apply early stopping, which is guided by two key validation metrics: the average approximation ratio (AR) over a held-out set of graphs and the circuit error rate, defined as the fraction of structurally invalid circuits generated by the model. A circuit is considered to be invalid if it does not respect the expected quantum circuit structure. Training terminates once these metrics converge or begin to degrade. 

\subsection*{Proposed Use Case for \sysname}
\begin{figure}
    \centering
    \includegraphics[
        width=1\linewidth,
    ]{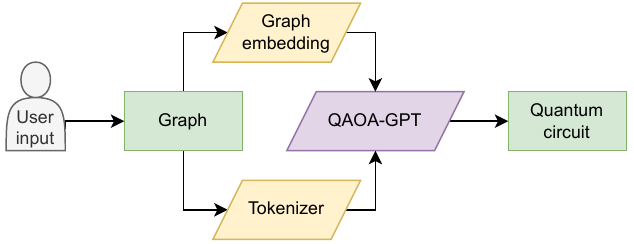}
    \caption{
      \label{fig:usecase_diagram}
      Proposed use case diagram. Given a user-supplied input graph, the system computes a fixed-length graph embedding and tokenizes the graph structure. Both representations are passed to the QAOA-GPT model, which autoregressively generates a quantum circuit that solves the corresponding QAOA optimization problem.
    }
\end{figure}

Once the model is fully trained, it is ready to generate quantum circuits for new, unseen graphs (Figure~\ref{fig:usecase_diagram}). They can be supplied as edgelists, which are then automatically embedded, tokenized and sent to the GPT model. As a result, we obtain quantum circuits generated specifically to give QAOA or QAOA-like solutions for given inputs, that are ready to be simulated classically or transpiled to be run on a quantum device.  

\section{Computational Results}

\begin{figure}
    \centering
    \includegraphics[
        width=1\linewidth,
    ]{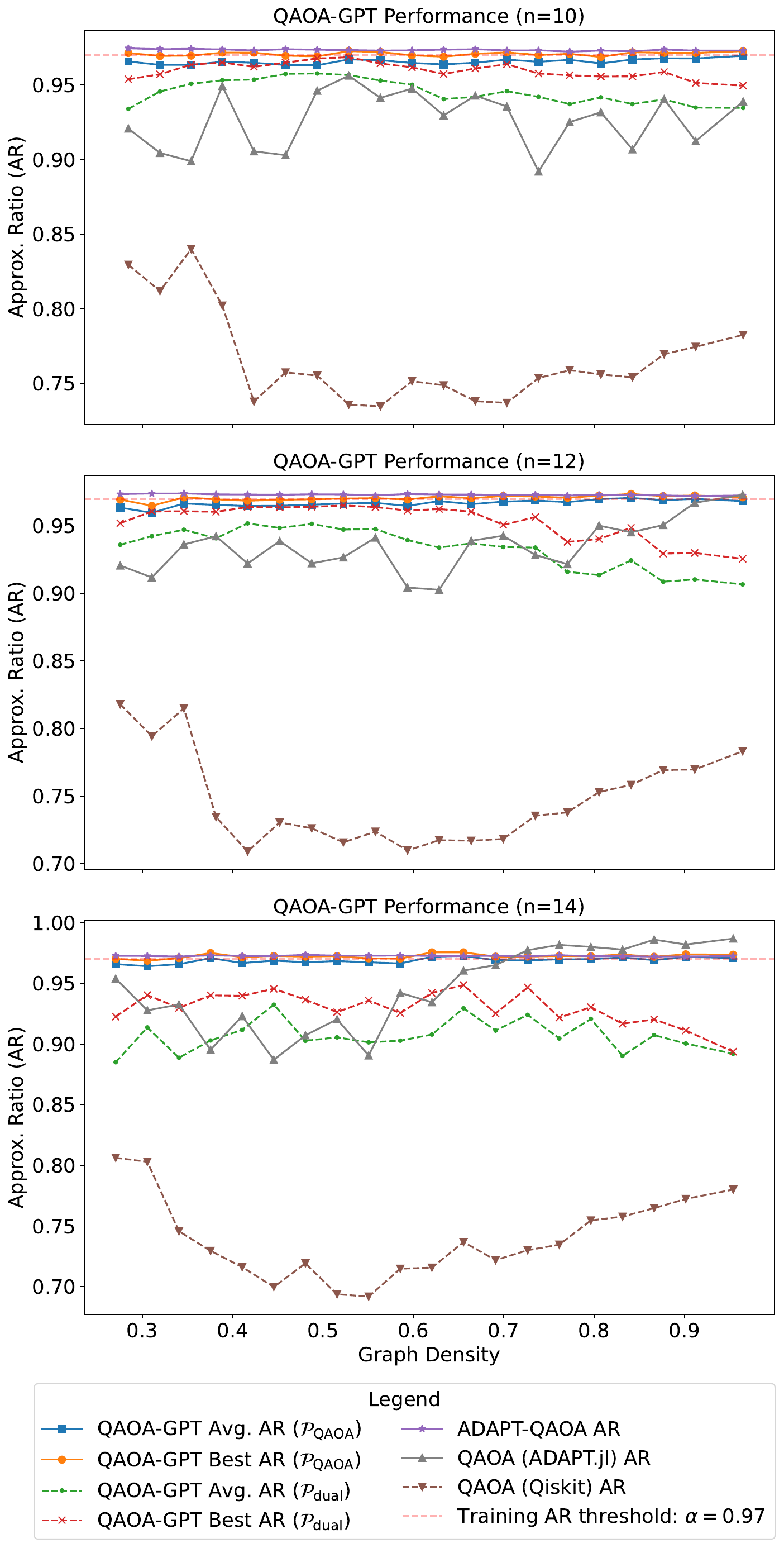}
    \caption{
      \label{fig:ar_stacked}
    Average approximation ratio (AR) of quantum circuits generated with different QAOA methods for graphs with \( n = 10, 12, 14 \) nodes across varying densities. Each point represents the mean AR over 50 weighted Erdős–Rényi graphs within a density bin. A total of 1000 random graphs were evaluated for each problem size. Solid curves show mainline methods: QAOA-GPT trained with \( \mathcal{P}_{\text{QAOA}} \), ADAPT-QAOA, and standard QAOA (ADAPT.jl). Dashed curves show systems that are included for comparison: QAOA-GPT trained with \( \mathcal{P}_{\text{dual}} \) (Eq.~\ref{eq:p_dual_op}) and Qiskit QAOA. Both average and best (out of 5 runs) ARs are reported for QAOA-GPT models. The horizontal dashed line marks the training threshold \( \alpha = 0.97 \). 
}
\end{figure}

\subsection{Training Dataset Generation}

For all experiments, for every problem size of $n$ nodes we generate circuits as described in Section \ref{sec:proposed_method} for over 50{,}000 Erdős–Rényi graphs with edge probabilities \( s \in \{0.3, 0.4, \ldots, 0.9\} \) uniformly distributed across the specified range and edge weights sampled from \( \mathcal{U}(0,1) \). Each graph was solved with ADAPT-QAOA multiple times, namely, for every graph, a grid search was performed over \( \gamma_0 \in \{0.01, 0.1, 0.5, 1\}\) initialization values. The approximation ratio threshold for circuits forming the training set is set to \( \alpha \geq 0.97 \).

The operator pool used to report results is \( \mathcal{P}_{\text{QAOA}} = \left\{ H_B \right\} \), where $H_B$ is QAOA Mixing Hamiltonian.
In addition, we train QAOA-GPT with ADAPT-QAOA dual-qubit operator pool, which is defined in~\cite{zhu2022adaptive}:

\begin{equation}
    \mathcal{P}_{\text{dual}} = \{ B_i C_j \mid B,C \in \{X,Y,Z\} \} \cup \mathcal{P}_{\text{single}}
    \label{eq:p_dual_op}
\end{equation}
where \( B_i C_j \) denotes all possible two-qubit Pauli products, \( \mathcal{P}_{\text{single}} = \{ X_i, Y_i \} \cup \{ \sum_i Y_i \} \cup \mathcal{P}_{\text{QAOA}} \).

To generate the training dataset, we utilize the  high-performance computing  cluster, where compute nodes are equipped with AMD EPYC 7502 32-core processors. Depending on the graph size and circuit depth, dataset generation may take from several hours to multiple weeks (see Circuit Generation and Quantum Simulation Scalability discussion), as we typically employ $500$ CPU workers in parallel. Used ADAPT-QAOA implementation is single-threaded and parallelism is achieved at the process level. For the GPT model training, we use a single NVIDIA A100 GPU node.

\subsection{QAOA-GPT, QAOA, and ADAPT-QAOA Circuit Generation Comparison}
\begin{table*}[!t]
  \centering
  \caption{
    Approximation Ratio (AR) and number of layers $p$ for \sysname, ADAPT-QAOA and regular QAOA. Results are presented as mean values across 1000 graphs for each problem size ± standard deviation. 
    }
        \begin{tabular}{ccccc|cccc}
        \toprule
         & \multicolumn{4}{c}{AR} & \multicolumn{4}{c}{Number of Layers $p$} \\
         Problem & QAOA-GPT & QAOA-GPT & ADAPT-QAOA & QAOA & QAOA-GPT & QAOA-GPT & ADAPT-QAOA & QAOA \\
        Size & (Best) &  &  &  & (Best) &  &  &  \\
        \hline
        10 & 0.971 ± 0.009 & 0.966 ± 0.011 & 0.974 ± 0.003 & 0.926 ± 0.108 & 9.237 ± 1.049 & 8.800 ± 1.048 & 8.254 ± 1.488 & 10 \\
        12 & 0.971 ± 0.009 & 0.967 ± 0.011 & 0.973 ± 0.002 & 0.934 ± 0.108 & 10.176 ± 1.319 & 9.764 ± 1.295 & 8.933 ± 1.872 & 12 \\
        14 & 0.972 ± 0.007 & 0.969 ± 0.009 & 0.973 ± 0.002 & 0.946 ± 0.102 & 10.697 ± 0.998 & 10.233 ± 1.008 & 9.404 ± 1.875 & 14 \\
        \hline
        \end{tabular}
  \label{tab:perf_comp_table}
\end{table*}

We report the performance of our framework using average approximation ratios (ARs) across graph instances with $n = 10, 12, 14$ vertices, stratified by graph density $|E|/ {|V| \choose 2}$ (Figure~\ref{fig:ar_stacked}). Each data point represents the mean AR over 50 random graphs sampled within the corresponding density bin. A total of 1000 test graphs were generated for evaluation of each problem size using the Erd\H{o}s--R\'enyi model (as described in the Section \ref{sec:proposed_method}), with edge weights drawn from the uniform distribution $\mathcal{U}(0,1)$. Aggregated results are present in Table~\ref{tab:perf_comp_table}. 

To evaluate QAOA-GPT, we generate 5 circuits for every test graph and report both the average and best AR achieved depicted as blue and orange curves, respectively. For comparison, we include results from the ADAPT-QAOA circuits used in training, as well as circuits generated with standard QAOA. According to the experiments with the predictive model in \cite{shaydulin2021classical}, the required QAOA depth rarely exceeds the number of nodes for the MaxCut problem on ER graphs, therefore we set the depth of QAOA circuits equal to problem size $p = n$. We use two QAOA implementations: ADAPT.jl\footnote{\url{https://github.com/kmsherbertvt/ADAPT.jl}} and Qiskit\footnote{\url{https://github.com/qiskit-community/qiskit-optimization}}.For ADAPT.jl QAOA implementation, the initial parameter $\gamma_0$ was set to $0.01$, following the recommendations in~\cite{zhu2022adaptive}. Note that standard QAOA does not include any stopping criteria beyond optimizer convergence; as a result, in some instances it exceeds the training AR threshold used for QAOA-GPT data generation. In case of ADAPT-QAOA, the AR is evaluated at every subsequent layer addition and the circuit generation terminates as soon as the threshold is passed, which results in more compact quantum circuits.

As shown in the plots, QAOA-GPT achieves consistently strong performance, closely matching the baseline implementations across all tested problem sizes and densities, when trained with $\mathcal{P}_{\text{QAOA}}$. Using dual-qubit operator pool slightly decreases the approximation ratio for the reasons we discuss in Section~\ref{sec:op_choice}.

\section{Discussion}

In the this section, we discuss multiple important aspects about the proposed work, perform ablation studies and outline the directions for our future work.

\subsection{Role of Graph Embeddings}
\begin{table}[!t]
  \centering
  \caption{
    Role of Graph Embeddings on Model Performance.
    Columns represent different random graph generators. Bold font indicates best results.
    }
        \begin{tabular}{llllll}
        \toprule
         & BA & BP & ER & REG & WS \\
        Model &  &  &  &  &  \\
        \midrule
        ER without FEATHER & 0.921 & 0.9 & 0.966 & 0.97 & \textbf{0.967} \\
        ER with FEATHER & \textbf{0.94} & \textbf{0.956} & \textbf{0.969} & \textbf{0.971} & \textbf{0.967} \\
        \bottomrule
        \end{tabular}
  \label{tab:emb_experiment}
\end{table}

In many prior studies, the performance of QAOA has been closely tied to the structural properties of the input graph. For example, graph low dimensional representation was used for parameter transferability between instances of QAOA \cite{falla2024graph}. The node degree parity and subgraph similarity were also helpful for parameter transferability \cite{galda2023similarity}. The classical graph structural symmetry properties were used to predict QAOA depth \cite{shaydulin2021classical}. Motivated by this, we incorporate low-dimensional graph representations generated by FEATHER as features in our quantum circuit generation pipeline. These embeddings capture essential structural information in a compact form, enabling the GPT model to condition circuit generation on problem-specific characteristics.

To evaluate the importance of graph embeddings in the \sysname~ framework, we conduct an ablation study in which two models are trained under identical conditions—with and without graph embeddings integrated into the architecture. Both models are trained using graphs induced by the Erdős–Rényi (ER) generator with $n=10$ nodes. 

We assess model generalization by testing on previously unseen graphs that are either in-distribution (ER graphs) or out-of-distribution, drawn from four other random graph generators: Barabási–Albert (BA), Bipartite (BP), Regular (REG), and Watts–Strogatz (WS) implemented in \cite{Hagberg08}. 

Table~\ref{tab:emb_experiment} reports the best approximation ratios, averaged over 300 graphs for each generator. The results demonstrate that incorporating graph embeddings substantially improves model performance across all graph types. Notably, the model without embeddings performs significantly worse on BA and BP graphs, suggesting that their structural complexity poses challenges when represented solely as edge lists, without any additional structural encoding.

These results support our design choice described in Section~\ref{sec:proposed_method}, where we integrate FEATHER-based graph embeddings into the transformer block input to enhance circuit generation capabilities.

\subsection{Training Data Distribution}
\begin{table}[!t]
  \centering
  \caption{
    QAOA-GPT performance based on the training set data variability. Bold font indicates best results.
    }
    \setlength{\tabcolsep}{.35em}
    \renewcommand{\arraystretch}{1.1}
        \begin{tabular}{lrrr|rrr}
        \toprule
        Graph Type & \multicolumn{3}{c}{BA} & \multicolumn{3}{c}{ER} \\
         & Best AR & Avg. AR & Err. Rate & Best AR & Avg. AR & Err. Rate \\
        Model &  &  &  &  &  &  \\
        \hline
        ER  & 0.94 & 0.925 & 0.093 & 0.969 & 0.96 & 0.01 \\
        ER \& BA  & \textbf{0.966} & \textbf{0.927} & 0 & \textbf{0.97} & 0.96 & 0 \\
        \bottomrule
        \end{tabular}
  \label{tab:ba_experiment}
\end{table}

To evaluate the effect of training data diversity on model generalization, we perform an additional experiment where two models are trained with graph embeddings: one using only Erdős–Rényi (ER) graphs, and the other using a mixture of ER and Barabási–Albert (BA) graphs. We then compare their performance on both ER and BA test sets (Table~\ref{tab:ba_experiment}).

The results indicate that incorporating BA graphs into the training set leads to a noticeable improvement in performance on BA graphs, particularly in terms of best and average approximation ratio, while maintaining comparable performance on ER graphs. This finding shows the importance of training on structurally diverse graphs to ensure robust generalization across different graph families.

\subsection{Training Dataset Target Quality}
\begin{figure}
    \centering
    \includegraphics[
        width=0.98\linewidth,
    ]{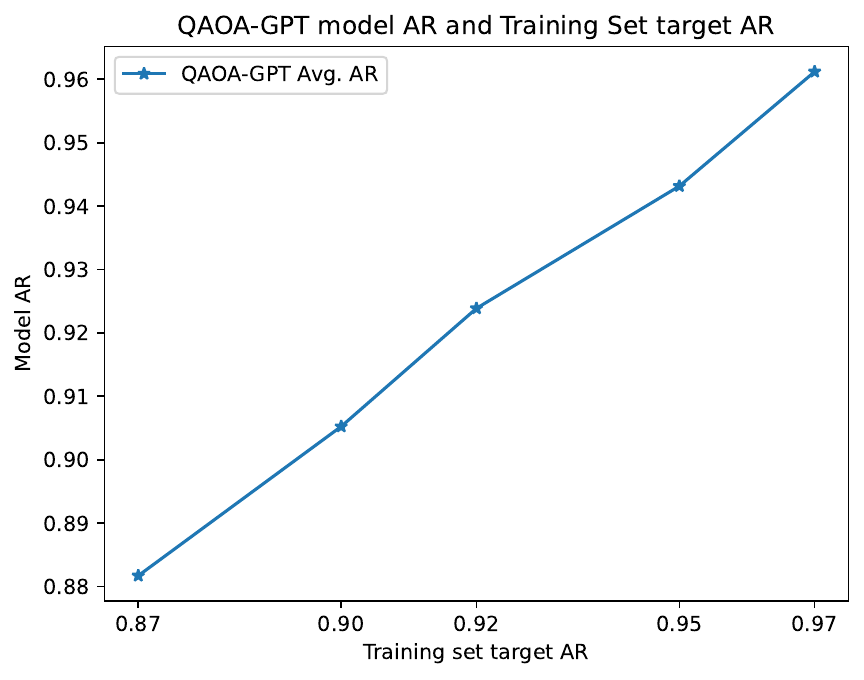}
    \caption{
      \label{fig:training_ar_target}
      \sysname~ performance as a function of the target approximation ratio (AR) used during training data generation. The experiment is conducted on graphs with $n = 8$ nodes. Each point represents the average AR achieved on a validation set by a separate~\sysname~ instance trained with circuits filtered by a given target AR threshold.
    }
\end{figure}

Another critical factor influencing model performance is the quality of the training dataset. In Figure~\ref{fig:training_ar_target}, we analyze the impact of the target approximation ratio (AR) threshold used during dataset generation on the final model accuracy. All experiments are conducted on graphs with $n = 8$ nodes.

The results reveal a clear connection between the target AR of the training circuits and the resulting performance of the model. As the quality of the training circuits increases, so does the average approximation ratio achieved by QAOA-GPT. This highlights the importance of including high-quality quantum circuits when constructing training datasets, as better training data directly translates to improved model performance.

\subsection{Training Progress}
\begin{figure}
    \centering
    \includegraphics[
        width=0.98\linewidth,
    ]{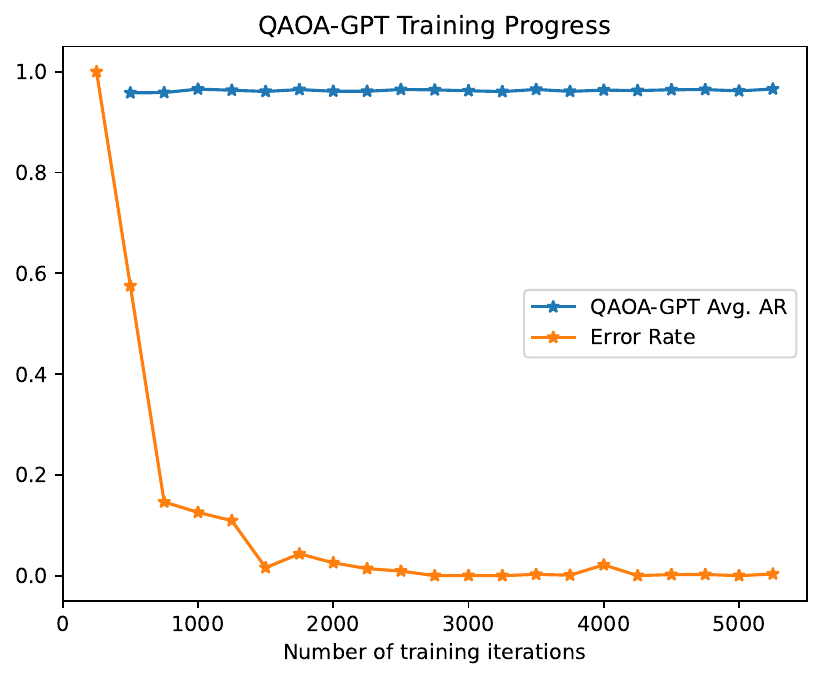}
    \caption{
      \label{fig:training_progress}
      \sysname~ performance over the course of training . The blue curve shows the average AR, and the orange curve shows the fraction of structurally invalid circuits.
    }
\end{figure}

Another important aspect to consider is the model's learning dynamics as it is exposed to increasing amounts of training data. In Figure~\ref{fig:training_progress}, we illustrate how the average approximation ratio and error rate of \sysname~ evolve over training iterations, measured on a held-out validation set.

Interestingly, the approximation ratio remains relatively stable throughout training, while the error rate decreases steadily. This behavior suggests that the model quickly learns to generate structurally valid and high-performing circuits early in training. Since the training data only includes circuits with high approximation ratios, the model is not exposed to examples of structurally valid but low-quality circuits. We can speculate that it has little opportunity to learn how to generate such circuits.

We also note that the training loss is purely autoregressive and does not directly incorporate energy-based or performance-aware terms. This may explain why the approximation ratio does not improve significantly during training, as long as structural validity is maintained.

Incorporating energy-sensitive training objectives, such as reinforcement learning with energy-based reward functions or Direct Preference Optimization (DPO) guided by approximation ratios, could provide more explicit supervision for high-quality circuit generation~\cite{minami2025generative}. We leave this direction for future work, as it may further improve the alignment between model predictions and quantum circuit performance.

\subsection{Remarks on Operator Pool Choice}
\label{sec:op_choice}
\begin{figure}
    \centering
    \includegraphics[
        width=0.98\linewidth,
    ]{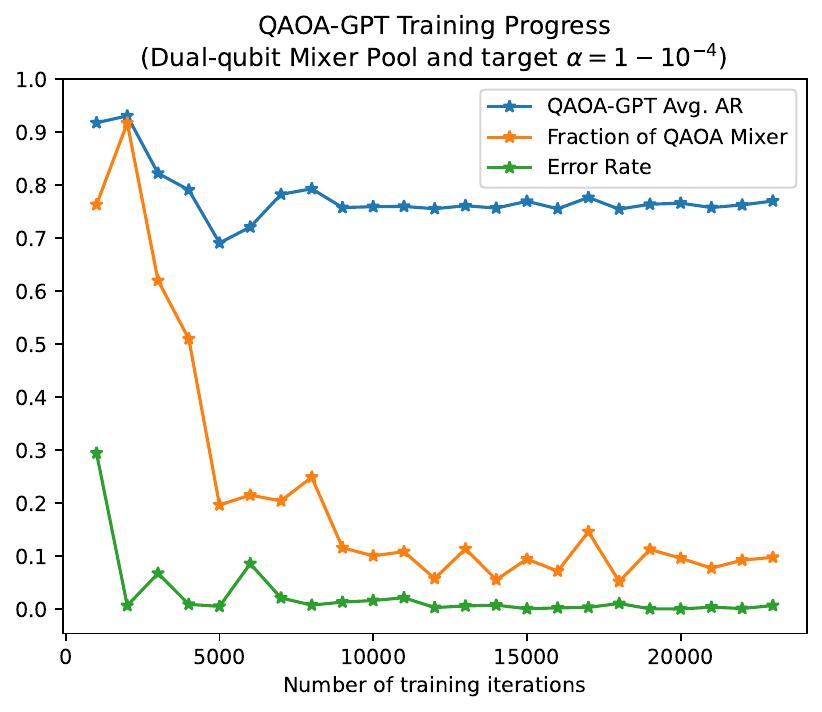}
    \caption{
      \label{fig:training_progress_high_ar}
      \sysname~ performance over the course of training on graphs with 6 nodes with high target AR and $\mathcal{P}_{\text{dual}} $ mixer pool. One can observe that the fraction of QAOA mixers is dropping as the number of training iterations is growing.
    }
\end{figure}

Operator pool selection significantly impacts circuit diversity, but its influence on circuit quality depends on the desired approximation ratio (AR). For our standard target threshold (\( \alpha \geq 0.97 \)), increasing the expressiveness of the operator pool has negligible effect on the final AR, consistent with findings in the original ADAPT-QAOA work~\cite{zhu2022adaptive}. However, when stricter energy tolerances are required (e.g., \( \alpha \geq 1 - 10^{-4} \)), the operator pool diversity becomes essential. Achieving such precision demands deeper and more complex circuits, which are not only harder to simulate but also harder to learn with traditional sequence models.

To investigate this, we train \sysname~ on a dataset generated using the ADAPT Dual-qubit mixer pool (Equation~\ref{eq:p_dual_op}).

We set the AR threshold to \( \alpha \geq 1 - 10^{-4} \) and problem size fixed at \( n = 6 \). Figure~\ref{fig:training_progress_high_ar} shows three training curves: (1) average approximation ratio, (2) error rate, and (3) the fraction of selected operators that are standard QAOA mixers.

Early in training, \sysname~ heavily favors the standard QAOA mixer, achieving relatively high AR values ($>0.9$) by learning the simplest successful strategy. As training progresses, the model begins to incorporate a more diverse set of operators reflecting patterns from the ADAPT Dual-qubit mixer pool, but the AR plateaus at approximately 0.8. This suggests that while the model adapts its operator distribution, it struggles to capture the structures required for high precision optimization.

These observations emphasize the need for better inductive biases, richer representations of operator semantics or energy feedback in future versions of \sysname. Structural embeddings of Pauli strings or attention mechanisms biased by operator commutativity could offer a promising direction.

\subsection{Circuit Generation Scalability}
\begin{figure}
    \centering
    \includegraphics[
        width=0.98\linewidth,
    ]{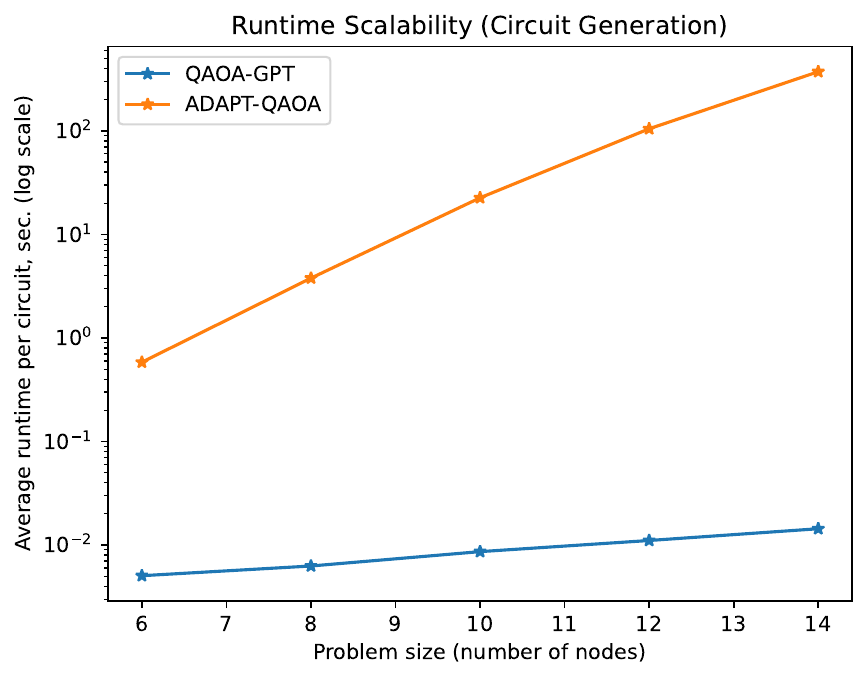}
    \caption{
      \label{fig:circ_gen_scal}
      Running time scalability with problem size. The plot shows that QAOA-GPT maintains nearly constant inference time due to its non-gradient-based generation, while ADAPT-QAOA runtime grows rapidly with problem size.
    }
\end{figure}

One significant aspect of our system is its fast inference runtime. The trained model is compact (around 12M parameters) and fits entirely on a single GPU. In Figure~\ref{fig:circ_gen_scal}, we present a runtime scalability plot comparing \sysname~ with ADAPT-QAOA framework used to generate our training data. The vertical axis is logarithmic to better visualize growth trends. Since \sysname~ does not perform gradient-based optimization during inference to select the operator pool to append to the circuit or optimize the circuit parameters, its runtime remains nearly constant as problem size increases, unlike the exponential scaling observed in classical gradient-based methods. While larger graphs result in longer token sequences (both for edge lists and generated circuits), this growth has only a minor impact on runtime for token-based circuit generation. \sysname~ supports efficient batching, which allows us to generate multiple circuits at a time, bringing average runtime per circuit even lower. 

\subsection{Quantum Simulation Scalability}
\begin{figure}
    \centering
    \includegraphics[
        width=0.98\linewidth,
    ]{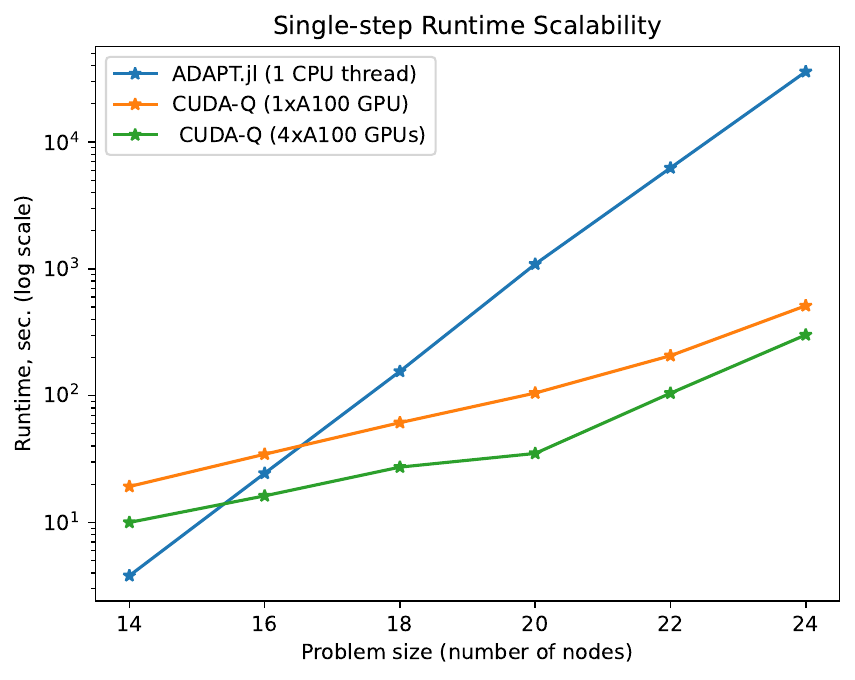}
    \caption{
      \label{fig:first_layer_scal} Time elapsed (in log scale) to compute the the energy gradient with respect to $\mathcal{P}_{\text{dual}} $ operator pool (Equation~\ref{eq:p_dual_op}) in the ADAPT-QAOA algorithm during the initial step where the reference state is $|+\rangle^{\otimes n}$. Simulations are performed for different problem size. The noiseless state vector simulator is employed for all simulations. The execution time of circuit to compute the gradient on single CPU using ADAPT-QAOA implemented in Julia (blue) is compared against ADAPT-QAOA implemented in CUDA-Q platform with GPU accelerated statevector: orange (single GPU) and green (four GPUs).}

\end{figure}

It is important to address the scalability and computational cost associated with training dataset generation. While \sysname~ demonstrates exceptional inference speed and circumvents the scalability limitations of gradient-based methods, its training data is nonetheless generated using such methods. To accelerate the dataset generation, in particular for large problem sizes, it is essential to use the cutting-edge software optimized to run on NVIDIA GPU-based platforms. In this work, we used 
the GPU accelerated 
 statevector simulator, which is a part of CUDA-Q~\cite{bayraktar2023cuquantum, cudaq_github} platform. CUDA-Q is an open-source quantum platform developed by NVIDIA Corporation~\cite{cuda_q, cudaq_github} and it is designed for programming tightly integrated heterogeneous quantum--GPU systems, like DGX Quantum~\cite{dgx_quantum}.

To assess the computational cost of this data generation, we measure the runtime of a single ADAPT-QAOA step, specifically, the elapsed time for the energy gradient evaluation with respect to  each operator from $\mathcal{P}_{\text{dual}} $ operator pool during the initial iteration of the ADAPT-QAOA. This is known to be the most expensive step in the ADAPT-QAOA approach. For every problem size, we use complete graphs to simulate the worst case scinario. Figure ~\ref{fig:first_layer_scal} presents the obtained runtime from ADAPT-QAOA implementation with Julia and CUDA-Q. In both cases, we use noisless statevector simulator. However, in the former, the statevctor is executed on CPU (AMD EPYC 7502 32-core processor), while in the latter we employ the GPU accelerated statevector simulator available in the CUDA-Q backend~\cite{bayraktar2023cuquantum}. NVIDIA A100 GPUs on NERSC's Perlmutter supercomputer is used to generate the data on GPU. Figure ~\ref{fig:first_layer_scal} shows that CUDA-Q on a single GPU achieves $~70\times$ speedup compared to Julia code on a single CPU for a graph with 24 nodes. Furthermore, by distributing the gradient evaluation across four GPUs, we observe $~1.69\times$ speedup compared to a single GPU with CUDA-Q.

In this experiment, we limited our problem size to 24 nodes, but this limit is not fundamental. As shown in the Figure~\ref{fig:first_layer_scal}, employing  GPU accelerated statevector simulator significantly improves scalability relative to CPU-based implementations, suggesting that generation of much larger training datasets is feasible and can be efficiently parallelized on GPUs using CUDA-Q platform. It is an important conclusion given that to achieve the full potential of GPT, it often requires the generation of a massive training dataset, which could be severely limited by the cost of ADAPT if GPUs are not used.

\section{Related work}

Two recent generative GPT-based quantum circuit generation approaches are the Generative Quantum Eigensolver and GroverGPT. The Generative Quantum Eigensolver (GQE)~\cite{nakaji2024generative} employs a decoder-only transformer to generate quantum circuits as sequences over a fixed operator pool, trained via logit matching against measured energy expectations. Unlike iterative ansatz construction in ADAPT-QAOA, GQE samples full circuits conditioned on prior quantum evaluations, enabling direct circuit synthesis and transfer across related Hamiltonians.

The GroverGPT~\cite{wang2024grovergpt} is an 8B-parameter LLaMA-based model fine-tuned to approximate Grover's quantum search algorithm. By learning from 97K quantum search instances, it achieved near-perfect accuracy on 6- and 10-qubit tasks and generalized to over 20 qubits with $>95\%$ accuracy, outperforming GPT-4o. Its architecture avoids explicit state-vector simulation by leveraging pattern recognition and quantum-aware prompting strategies.

Layer VQE~\cite{liu2022layer} in another example of adaptive approach. It introduces a layerwise ansatz growth strategy inspired by ADAPT-VQE, where circuit depth is incrementally increased without gradient-based operator selection. Unlike ADAPT-QAOA, which adaptively constructs circuits using operator pools and energy gradients, Layer VQE grows a predefined hardware-efficient ansatz, trading fine-grained adaptivity for simplicity and robustness to sampling noise.

\section{Conclusion}

We have presented QAOA-GPT, a generative AI framework that employs transformer--based models to automatically synthesize compact quantum circuits for solving combinatorial optimization problems using QAOA and QAOA-like strategies. Traditional QAOA relies on iterative classical optimization over variational parameters, a process that is often computationally intensive, sensitive to initialization, and prone to getting trapped in local minima—especially as circuit depth and problem size increase. By replacing this iterative optimization loop with a single-pass GPT-based inference model, QAOA-GPT circumvents these challenges, significantly reducing the time required to generate compact quantum circuits. Due to the limitations in the quantum device capacity such accelerations are vital as the global optimization problem is often decomposed and solved by parts \cite{bach2024mlqaoa,cameron2024scaling}.

Our framework builds on the demonstrated success of GPT-based code generation in classical domains, where models like Codex and Code Llama have shown exceptional ability to synthesize and generalize across a wide range of programming tasks. We extend this paradigm to quantum computing, where quantum circuits can be viewed as structured code sequences. Leveraging GPT’s strength in modeling such sequences, QAOA-GPT learns to generate complete circuit solutions conditioned on problem-specific representations, effectively serving as a scalable, data-driven alternative to handcrafted or adaptive ansatz construction.

For many combinatorial optimization problems, exact solutions are not required, and near-optimal results are often sufficient in practice. In our main experiments, we generated training data consisting of circuits that achieved at least 97\% of the optimal solution quality. We observed that for smaller graphs and this level of approximation, only modest training was needed to achieve strong performance. However, as problem sizes or circuit depths increase and specific applications demand higher approximation ratios, more extensive training will be necessary to capture the richer structural patterns and circuit configurations required. This is justified by the increasing complexity of the solution space, which necessitates a larger and more expressive model to generalize effectively. Nevertheless, we advocate that even from smaller graph learning, we can try to understand the structure of circuits other than QAOA high-quality circuits, which is a promising future research direction.

The rapid advancement of generative AI, like GPT models, together with GPU-accelerated platforms developed by NVIDIA, is fueling a technological shift that is redefining the landscape of both classical and quantum computation. By integrating AI capabilities, we can address key challenges in quantum computing, such as the efficient generation of compact, high-quality quantum circuits. This combination of large-scale AI models and GPU-accelerated quantum simulation is an excellent example of how AI is becoming a critical enabler of progress in quantum algorithm design.

\section*{Acknowledgment}
This work was partially supported with funding from the Defense Advanced Research Projects Agency (DARPA) under the ONISQ program and by National Science Foundation ExpandQISE program award \#2427042. K. Sherbert and K. Shirali were supported by the DOE Office of Science, National Quantum Information Science Research Centers, Co-design Center for Quantum Advantage (C2QA) under contract number DE-SC0012704. M.H.F. acknowledge using resources of the National Energy Research Scientific Computing Center, a DOE Office of Science User Facility supported by the Office of Science of the U.S. Department of Energy under Contract No. DE-AC02-05CH11231 using NERSC award NERSC DDR-ERCAP0033101

\bibliographystyle{unsrt}
\bibliography{ilya-biblio,bibliography}

\begin{thebibliography}{10}

\bibitem{herman2023quantum}
Dylan Herman, Cody Googin, Xiaoyuan Liu, Yue Sun, Alexey Galda, Ilya Safro, Marco Pistoia, and Yuri Alexeev.
\newblock Quantum computing for finance.
\newblock {\em Nature Reviews Physics}, 5(8):450--465, 2023.

\bibitem{cao2019quantum}
Yudong Cao, Jonathan Romero, Jonathan~P Olson, Matthias Degroote, Peter~D Johnson, M{\'a}ria Kieferov{\'a}, Ian~D Kivlichan, Tim Menke, Borja Peropadre, Nicolas~PD Sawaya, Sukin Sim, Libor Veis, and Alan Aspuru-Guzik.
\newblock Quantum chemistry in the age of quantum computing.
\newblock {\em Chemical reviews}, 119(19):10856--10915, 2019.

\bibitem{matsci_qc}
Yuri Alexeev, Maximilian Amsler, Marco~Antonio Barroca, Sanzio Bassini, Torey Battelle, Daan Camps, and et~al.
\newblock Quantum-centric supercomputing for materials science: A perspective on challenges and future directions.
\newblock {\em Future Generation Computer Systems}, 160:666--710, 2024.

\bibitem{shaydulin2019hybrid}
Ruslan Shaydulin, Hayato Ushijima-Mwesigwa, Christian F.~A. Negre, Ilya Safro, Susan~M. Mniszewski, and Yuri Alexeev.
\newblock A hybrid approach for solving optimization problems on small quantum computers.
\newblock {\em Computer}, 52(6):18--26, June 2019.

\bibitem{biamonte2017quantum}
Jacob Biamonte, Peter Wittek, Nicola Pancotti, Patrick Rebentrost, Nathan Wiebe, and Seth Lloyd.
\newblock Quantum machine learning.
\newblock {\em Nature}, 549(7671):195--202, 2017.

\bibitem{farhi2014QAOA}
Edward Farhi, Jeffrey Goldstone, and Sam Gutmann.
\newblock A quantum approximate optimization algorithm.
\newblock {\em arXiv preprint arXiv:1411.4028}, 2014.

\bibitem{blekos2024review}
Kostas Blekos, Dean Brand, Andrea Ceschini, Chiao-Hui Chou, Rui-Hao Li, Komal Pandya, and Alessandro Summer.
\newblock A review on quantum approximate optimization algorithm and its variants.
\newblock {\em Physics Reports}, 1068:1--66, 2024.

\bibitem{zhu2022adaptive}
Linghua Zhu, Ho~Lun Tang, George~S Barron, FA~Calderon-Vargas, Nicholas~J Mayhall, Edwin Barnes, and Sophia~E Economou.
\newblock Adaptive quantum approximate optimization algorithm for solving combinatorial problems on a quantum computer.
\newblock {\em Physical Review Research}, 4(3):033029, 2022.

\bibitem{grimsley2019adaptive}
Harper~R Grimsley, Sophia~E Economou, Edwin Barnes, and Nicholas~J Mayhall.
\newblock An adaptive variational algorithm for exact molecular simulations on a quantum computer.
\newblock {\em Nature communications}, 10(1):1--9, 2019.

\bibitem{liu2022layer}
Xiaoyuan Liu, Anthony Angone, Ruslan Shaydulin, Ilya Safro, Yuri Alexeev, and Lukasz Cincio.
\newblock Layer {VQE}: A variational approach for combinatorial optimization on noisy quantum computers.
\newblock {\em {IEEE Transactions on Quantum Engineering}}, 3:1--20, 2022.

\bibitem{wang2024grovergpt}
Haoran Wang, Pingzhi Li, Min Chen, Jinglei Cheng, Junyu Liu, and Tianlong Chen.
\newblock Grovergpt: A large language model with 8 billion parameters for quantum searching.
\newblock {\em arXiv preprint arXiv:2501.00135}, 2024.

\bibitem{nakaji2024generative}
Kouhei Nakaji, Lasse~Bj{\o}rn Kristensen, Jorge~A Campos-Gonzalez-Angulo, Mohammad~Ghazi Vakili, Haozhe Huang, Mohsen Bagherimehrab, Christoph Gorgulla, FuTe Wong, Alex McCaskey, Jin-Sung Kim, et~al.
\newblock The generative quantum eigensolver (gqe) and its application for ground state search.
\newblock {\em arXiv preprint arXiv:2401.09253}, 2024.

\bibitem{bayraktar2023cuquantum}
Harun Bayraktar, Ali Charara, David Clark, Saul Cohen, Timothy Costa, Yao-Lung~L Fang, Yang Gao, Jack Guan, John Gunnels, Azzam Haidar, et~al.
\newblock cuquantum sdk: A high-performance library for accelerating quantum science.
\newblock In {\em 2023 IEEE International Conference on Quantum Computing and Engineering (QCE)}, volume~1, pages 1050--1061. IEEE, 2023.

\bibitem{cudaq_github}
{NVIDIA Corporation}.
\newblock {NVIDIA CUDA-Q framework}.
\newblock \url{https://github.com/NVIDIA/cuda-quantum}.

\bibitem{NP_completeQAOA}
Michael~R Garey, David~S Johnson, and Larry Stockmeyer.
\newblock Some simplified np-complete problems.
\newblock In {\em Proceedings of the sixth annual ACM symposium on Theory of computing}, pages 47--63, 1974.

\bibitem{glover2022quantum}
Fred Glover, Gary Kochenberger, Rick Hennig, and Yu~Du.
\newblock Quantum bridge analytics i: a tutorial on formulating and using qubo models.
\newblock {\em Annals of Operations Research}, 314(1):141--183, 2022.

\bibitem{glover2018tutorial}
Fred Glover and Gary Kochenberger.
\newblock A tutorial on formulating qubo models.
\newblock {\em arXiv preprint arXiv:1811.11538}, 2018.

\bibitem{ushijima2021multilevel}
Hayato Ushijima-Mwesigwa, Ruslan Shaydulin, Christian~FA Negre, Susan~M Mniszewski, Yuri Alexeev, and Ilya Safro.
\newblock Multilevel combinatorial optimization across quantum architectures.
\newblock {\em ACM Transactions on Quantum Computing}, 2(1):1--29, 2021.

\bibitem{bravyi2022hybrid}
Sergey Bravyi, Alexander Kliesch, Robert Koenig, and Eugene Tang.
\newblock Hybrid quantum-classical algorithms for approximate graph coloring.
\newblock {\em Quantum}, 6:678, 2022.

\bibitem{tsvelikhovskiy2024equivariant}
Boris Tsvelikhovskiy, Ilya Safro, and Yuri Alexeev.
\newblock Equivariant qaoa and the duel of the mixers.
\newblock {\em arXiv preprint arXiv:2405.07211}, 2024.

\bibitem{radford2018improving}
Alec Radford, Karthik Narasimhan, Tim Salimans, Ilya Sutskever, et~al.
\newblock Improving language understanding by generative pre-training.
\newblock 2018.

\bibitem{vaswani2017attention}
Ashish Vaswani, Noam Shazeer, Niki Parmar, Jakob Uszkoreit, Llion Jones, Aidan~N Gomez, {\L}ukasz Kaiser, and Illia Polosukhin.
\newblock Attention is all you need.
\newblock In {\em Advances in neural information processing systems}, pages 5998--6008, 2017.

\bibitem{chen2021evaluating}
Mark Chen, Jerry Tworek, Heewoo Jun, Qiming Yuan, Henrique Ponde De~Oliveira Pinto, Jared Kaplan, Harri Edwards, Yuri Burda, Nicholas Joseph, Greg Brockman, et~al.
\newblock Evaluating large language models trained on code.
\newblock {\em arXiv preprint arXiv:2107.03374}, 2021.

\bibitem{rozemberczki2020characteristic}
Benedek Rozemberczki and Rik Sarkar.
\newblock Characteristic functions on graphs: Birds of a feather, from statistical descriptors to parametric models.
\newblock In {\em Proceedings of the 29th ACM international conference on information \& knowledge management}, pages 1325--1334, 2020.

\bibitem{BURER2002}
Samuel Burer, Renato Monteiro, and Yin Zhang.
\newblock Rank-two relaxation heuristics for max-cut and other binary quadratic programs.
\newblock {\em SIAM Journal on Optimization}, 12, 07 2001.

\bibitem{shaydulin2021classical}
Ruslan Shaydulin, Stuart Hadfield, Tad Hogg, and Ilya Safro.
\newblock Classical symmetries and the quantum approximate optimization algorithm.
\newblock {\em Quantum Information Processing}, 20(11):1--28, 2021.

\bibitem{falla2024graph}
Jose Falla, Quinn Langfitt, Yuri Alexeev, and Ilya Safro.
\newblock Graph representation learning for parameter transferability in quantum approximate optimization algorithm.
\newblock {\em Quantum Machine Intelligence}, 6(2):46, 2024.

\bibitem{galda2023similarity}
Alexey Galda, Eesh Gupta, Jose Falla, Xiaoyuan Liu, Danylo Lykov, Yuri Alexeev, and Ilya Safro.
\newblock Similarity-based parameter transferability in the quantum approximate optimization algorithm.
\newblock {\em Frontiers in Quantum Science and Technology}, 2, 2023.

\bibitem{Hagberg08}
A.~Hagberg, D.A. Schult, and P.J. Swart.
\newblock Exploring network structure, dynamics, and function using networkx.
\newblock In {\em Proceedings of the 7th Python in Science Conference SciPy2008}, volume 836, pages 11--15, 2008.

\bibitem{minami2025generative}
Shunya Minami, Kouhei Nakaji, Yohichi Suzuki, Al{\'a}n Aspuru-Guzik, and Tadashi Kadowaki.
\newblock Generative quantum combinatorial optimization by means of a novel conditional generative quantum eigensolver.
\newblock {\em arXiv preprint arXiv:2501.16986}, 2025.

\bibitem{cuda_q}
{NVIDIA Corporation}.
\newblock {NVIDIA CUDA-Q}.
\newblock \url{https://developer.nvidia.com/cuda-q}, 2025.
\newblock Accessed: 2025-04-09.

\bibitem{dgx_quantum}
{NVIDIA Corporation}.
\newblock {NVIDIA DGX Quantum}.
\newblock \url{https://www.nvidia.com/en-us/data-center/dgx-quantum/}, 2025.
\newblock Accessed: 2025-04-09.

\bibitem{bach2024mlqaoa}
Bao Bach, Jose Falla, and Ilya Safro.
\newblock Mlqaoa: Graph learning accelerated hybrid quantum-classical multilevel qaoa.
\newblock In {\em 2024 IEEE International Conference on Quantum Computing and Engineering (QCE)}, volume~01, pages 1--12, 2024.

\bibitem{cameron2024scaling}
Cameron Ibrahim, Teague Tomesh, Zain Saleem, and Ilya Safro.
\newblock Scaling up the quantum divide and conquer algorithm for combinatorial optimization.
\newblock In {\em 2024 IEEE International Conference on Quantum Computing and Engineering (QCE)}, volume~1, pages 545--551. IEEE, 2024.
\newblock arXiv preprint arXiv:2405.00861.

\end{thebibliography}

\end{document}